# The Modified Scheme is still vulnerable to the parallel Session Attack

Manoj Kumar

Department of Mathematics, Rashtriya Kishan (P.G.) College

Shamli- Muzaffarnagar-247776

yamu_balyan@yahoo.co.in

*Abstract—* *In 2002, Chien–Jan–Tseng introduced an efficient remote user authentication scheme using smart cards. Further, in 2004, W. C. Ku and S. M. Chen proposed an efficient remote user authentication scheme using smart cards to solve the security problems of Chien et al.'s scheme. Recently, Hsu and Yoon et al. pointed out the security weakness of the Ku and Chen's scheme Furthermore, Yoon et al. modified the password change phase of Ku and Chen's scheme and they also proposed a new efficient remote user authentication scheme using smart cards. This paper analyzes that the modified scheme of Yoon et al. still vulnerable to parallel session attack.*

I. INTRODUCTION

To gain the access rights on an *authentication server* (*AS*) a password based remote user authentication schemes is used. The remote user makes a login request with the help of some secret information which are provided by the *AS*. On the other side the *AS* checks the validity of a login request made by a remote user *U*. In these schemes, the *AS* and the remote user *U* share a secret, which is often called as password. With the knowledge of this password, the remote user *U* uses it to create a valid login request to the *AS*. *AS* checks the validity of the login request to provide the access rights to the user *U*. Password authentication schemes with smart cards have a long history in the remote user authentication environment. So far different types of password authentication schemes with smarts cards [3] - [4] - [5] - [6] - [12] - [13] - [14] - [18] - [20] - [21] - [24] - [29] have been proposed.

Lamport [17] proposed the first well-known remote password authentication scheme using smart cards. In Lamport's scheme, the *AS* stores a password table at the server to check the validity of the login request made by the user. However, high hash overhead and the necessity for password resetting decrease the suitability and practical ability of Lamport's scheme. In addition, the Lamport scheme is vulnerable to a small *n* attack [7]. Since then, many similar schemes [23]-[26] have been proposed. They all have a common feature: *a verification password table should be securely stored in the AS*. Actually, this property is a disadvantage for the security point of view. If the password table is stolen /removed /modified by the adversary, the *AS* will be partially or totally braked/affected.

In 2002, Chien–Jan–Tseng [13] introduced an efficient remote user authentication scheme using smart cards. In 2004, Ku and Chen [31] pointed out some attacks [7]-[28]-[30] on Chien – Jan and Tseng's scheme. According to Ku and Chen, Chien et al.'s scheme is vulnerable to a reflection attack [7] and an insider attack [30]. Ku and Chen claimed that Chien et al.'s scheme is also not reparable [30]. In addition, they also proposed an improved scheme to prevent the attacks: reflection attack and an insider attack on Chien–Jan–Tseng's scheme. In the same year, Hsu [10] pointed out that the Chien–Jan–Tseng's scheme is still vulnerable to a parallel session attack and Yoon et al. [11] claimed that the password change phase of improved scheme of Chien–Jan–Tseng's scheme is still insecure.

### A. Contributions

This paper analyzes that the modified scheme of Yoon et al. is still vulnerable to parallel session attack.

### B. Organization

Section II reviews the Ku and Chen's scheme [31]. Section III reviews Hsu [10] and Yoon et al.'s comments on Ku and Chen's scheme .Section IV reviews Yoon et al.'s scheme [11]. Section V is about our observations on the security pitfall of Yoon et al.'s scheme. Finally, comes to a conclusion in the section VI.

## II. REVIEW OF KU AND CHEN'S SCHEME

This section briefly describes Ku and Chen's scheme [31]. This scheme has four phases: the registration phase, login phase, verification phase and the password change phase. All these four phases are described below.

### A. Registration Phase

This phase is invoked whenever $U$ initially or re-registers to $AS$. Let $n$ denotes the number of times $U$ re-registers to AS. The following steps are involved in this phase.

Step.R1: User $U$ selects a random number $b$ and computes $PW_S = f(b \oplus PW)$ and submits her/his identity $ID$ and $PW_S$ to the $AS$ through a secure channel.

Step.R2: $AS$ computes a secret number $R = f(EID \oplus x) \oplus PW_S$, where $EID = (ID \| n)$ and creates an entry for the user $U$ in his account database and stores $n = 0$ for initial registration, otherwise set $n = n+1$, and $n$ denotes the present registration.

Step.R3: $AS$ provides a smart card to the user $U$ through a secure channel. The smart card contains the secret number $R$ and a one-way function $f$.

Step.R4:  User $U$ enters his random number $b$ into his smart card.

### B. Login Phase

For login, the user $U$ inserts her/his smart card to the smart card reader and then keys the identity and the password to gain the access services. The smart card will perform the following operations:

Step.L1:  Computes $C_1 = R \oplus f(b \oplus PW)$ and $C_2 = f(C_1 \oplus T_U)$. Here $T_U$ denotes the current date and time of the smart card reader.

Step.L2:  Sends a login request $C = (ID, C_2, T_U)$ to the $AS$.

### C. Verification Phase

Assume $AS$ receives the message $C$ at time $T_S$, where $T_S$ is the current date and time at $AS$. Then the $AS$ takes the following actions:

Step.V1:  If the identity $ID$ and the time $T_U$ are not valid, then $AS$ will rejects this login request.

Step.V2:  Checks, if $C_2 \stackrel{?}{=} f(f(EID \oplus x) \oplus T_U)$, then the $AS$ accepts the login request and computes $C_3 = f(f(EID \oplus x) \oplus T_S)$. Otherwise, the login request $C$ will be rejected.

Step.V3:  $AS$ sends the pair $(C_3, T_S)$ to the user $U$ for mutual authentication.

Step.V4:  If the time $T_S$ is invalid i.e. $T_U = T_S$, then $U$ terminates the session. Otherwise, the user $U$ verifies the equation $C_3 \stackrel{?}{=} f(C_1 \oplus T_S)$ to authenticate $AS$.

### D. Password Change Phase

This phase is invoked whenever $U$ wants to change his password $PW$ with a new password, say $PW_{new}$. This phase has the following steps.

Step.P1: *U* inserts her/his smart card to the smart card reader keys the identity and the password and then requests to change the password. Next, *U* enters a new password $PW_{new}$.

Step.P2: *U*'s smart cards computes a new secret number $R_{new} = R \oplus f(b \oplus PW) \oplus f(b \oplus PW_{new})$ and then replaces $R$ with $R_{new}$.

## III. REVIEW OF HSU AND YOON ET AL.'S COMMENT ON THE KU AND CHEN'S SCHEME

### A. Hsu's Comment

According to Hsu, Ku and Chen's scheme is vulnerable to a parallel session attack [10]. The intruder Bob intercepts the communication between the *AS* and user *U* and then from this intercepted information, he makes a valid login request to masquerade as a legal user. The intruder Bob applies the following steps for a successful parallel session attack.

- ✓ Intercepts the login request $C = (ID, C_2, T_U)$ which is sent by a valid user *U* to *AS*.
- ✓ Intercepts the response message $(C_3, T_S)$, which is sent by *AS* to the user *U*.
- ✓ Starts a new session with the *AS* by sending a fabricated login request $C_f = (ID, C_3, T_S)$.

The fabricated login request passes all the requirements for a successful authentication of the intruder Bob by the *AS*, due to the fact that the second part, $C_3$, of the login request also satisfies the verification equation $C_3 \stackrel{?}{=} f(f(EID \oplus x) \oplus T_S)$.

- ✓ Finally, *AS* computes $C_4 = f(f(EID \oplus x) \oplus T_S)$ and responses with the message pair $(T^*_S, C_4)$ to the user *U* for mutual authentication, where is the current timestamp of the *AS*. Thus, the intruder intercepts and drops this message

*B. Yoon et al.'s Comment on Ku and Chen's Scheme*

According to Yoon et al., the password change phase of Ku and Chen's scheme is insecure. When the smart card was stolen, an authorized user can easily replace the old password by a new password of her/his choice. First, the authorized user inters the smart card into the smart card reader, enters the identity *ID* and any password $PW^*$ of her/his choice and then requests to change the password. Next, the authorized user enters a new password $PW^*_{new}$ and then the smart card computes a new $R^*_{new} = R \oplus f(b \oplus PW^*) \oplus f(b \oplus PW^*_{new})$ and then replaces $R$ with $R^*_{new}$, without any checking.

Thus, if the malicious user stole the user *U*'s smart card once, only for a small time and then change the valid password with an arbitrary password $PW^*$, then the registered/ legal user *U* also will not be able to make a valid login request. The *AS* will not authenticate a registered user *U*, because $C_2 \neq f(f(EID \oplus x) \oplus T_U)$ in the verification phase.

## IV. YOON ET AL.'S SCHEME

This section briefly describes Yoon et al.'s scheme [11]. This scheme also has four phases: the registration phase, login phase, verification phase and the password change phase. All these four phases are described below.

*A. Registration Phase*

This phase is invoked whenever *U* initially or re-registers to *AS*. Let *n* denotes the number of times *U* re-registers to AS. The following steps are involved in this phase.

- ❖ User *U* selects a random number *b* and computes $PW_S = f(b \oplus PW)$ and submits her/his identity *ID* and $PW_S$ to the *AS* through a secure channel.

- ❖ AS computes two secret numbers $V = f(EID \oplus x)$ and $R = f(EID \oplus x) \oplus PW_S$, where $EID = (ID \| n)$ and creates an entry for the user $U$ in his account database and stores $n = 0$ for initial registration, otherwise set $n = n+1$, and $n$ denotes the present registration.
- ❖ AS provides a smart card to the user $U$ through a secure channel. The smart card contains two secret numbers $V, R$ and a one-way function $f$.
- ❖ User $U$ enters her/his random number $b$ into his smart card.

B. *Login Phase*

For login, the user $U$ inserts her/his smart card to the smart card reader and then keys the identity and the password to gain access services. The smart card will perform the following operations:

- ❖ Computes $C_1 = R \oplus f(b \oplus PW)$ and $C_2 = f(C_1 \oplus T_U)$. Here $T_U$ denotes the current date and time of the smart card reader.
- ❖ Sends a login request $C = (ID, C_2, T_U)$ to the AS.

C. *Verification Phase*

Assume AS receives the message $C$ at time $T_S$, where $T_S$ is the current date and time at AS. Then the AS takes the following actions:

- ❖ If the identity $ID$ and the time $T_U$ is invalid i.e. $T_U = T_S$, then AS will rejects this login request.
- ❖ Checks, if $C_2 \stackrel{?}{=} f(f(EID \oplus x) \oplus T_U)$, then the AS accepts the login request and computes $C_3 = f(f(EID \oplus x) \oplus T_S)$. Otherwise, the login request $C$ will be rejected.
- ❖ AS sends the pair $(C_3, T_S)$ to the user $U$ for mutual authentication.
- ❖ If the time $T_S$ is invalid i.e. $T_U = T_S$, then $U$ terminates the session. Otherwise, $U$ verifies the equation $C_3 \stackrel{?}{=} f(C_1 \oplus T_S)$ to authenticates AS.

### D. Password Change Phase

This phase is invoked whenever $U$ wants to change his password $PW$ with a new one, say $PW_{new}$. This phase has the following steps.

- ❖ $U$ inserts her/his smart card to the smart card reader and then keys her/his identity and the old password $PW$ and then requests to change the password.
- ❖ $U$'s smart cards computes $V^* = R \oplus f(b \oplus PW)$.
- ❖ Compare this calculated value $V^*$ with the secret value $V$, which is stored in the smart card of the user $U$. If they are equal, then $U$ can select a new password $PW_{new}$, otherwise the smart card rejects the password change request.
- ❖ $U$'s smart cards computes a new secret number $R_{new} = V^* \oplus f(b \oplus PW_{new})$ and then replaces $R$ with $R_{new}$.

## V. OUR OBSERVATION: PARALLEL SESSION ATTACK ON YOON ET AL.'S SCHEME

Although, Yoon et al. [11] modified Ku and Chen's scheme to remove its security weaknesses against parallel session attack. But, we analyze that the modified scheme of Yoon et al. is still vulnerable parallel session attack. This following section proves our claim that the modified scheme is still vulnerable a parallel session attack by an intruder.

Since, a remote user password authentication is used to authenticate the legitimacy of the remote users over an insecure channel. Thus, an intruder Bob is able to intercept all the communication between the *AS* and user $U$ and then from this intercepted information, he makes a valid login request to masquerade as a legal user. The intruder Bob applies the following steps for a successful parallel session attack on Yoon et al.'s scheme.

- ✓ Intercepts the login request $C = (ID, C_2, T_U)$ which is sent by a valid user $U$ to $AS$. In this login request $C$, the time $T_U$ is the current time of the smart card reader, whenever the user $U$ makes the login request.
- ✓ Intercepts the response message $(C_3, T_S)$, which is sent by $AS$ to he user $U$. In this response message, the time $T_S$ in the current time at the $AS$, when $AS$ receives the login request $C$.
- ✓ Starts a new session with the $AS$ by sending a fabricated login request $C_f = (ID, C_3, T_S)$.

Upon, receiving the fabricated login request $C_f = (ID, C_3, T_S)$, at time $T_S^*$, where $T_S^*$ is the current date and time at $AS$. The $AS$ performs the following steps to ensure the validity of the received login request.

- ❖ Checks the validity of the format of the identity $ID$ and the time $T_U$ i.e. $T_S^* \neq T_S$. Both these conditions hold true, because the intruder has been used a previously registered identity $ID$ and obviously the time $T_S^*$ will be different from the time $T_S$.
- ❖ Checks, the verification equation $C_3 \stackrel{?}{=} f(f(EID \oplus x) \oplus T_S)$, which is also holds truly. The logic behind the successful verification of this phase is very interesting. If we observe login and verification phase of Yoon et al.'s scheme, then it makes a sense that the second part $C_2$ of the login request $C = (ID, C_2, T_U)$ and the first part $C_3$ of the response message $(C_3, T_S)$ are computed by the same procedure and with similar information.
- ❖ $AS$ sends the pair $T_S$ and $C_3$ to the user $U$ for mutual authentication.
- ❖ If the time $T_S$ is invalid i.e. $T_U = T_S$, then $U$ terminates the session. Otherwise, $U$ verifies the equation $C_3 \stackrel{?}{=} f(C_1 \oplus T_S)$ to authenticate $AS$.

❖ Finally, *AS* computes $C_4 = f(f(EID \oplus x) \oplus T^*_S)$ and responses with the message pair $(C_4, T^*_S)$ to the user *U* for mutual authentication, where $T^*_S$ is the current timestamp of the *AS*. Thus, the intruder intercepts and drops this message

In this way, the fabricated login request $C_f = (ID, C_3, T_S)$, which is made by the intruder, satisfies the all the requirements for a successful authentication of the intruder Bob by the *AS*.

## VI. CONCLUSION

This paper analyzed the modified scheme of Yoon et al.'s scheme is still vulnerable to the parallel session attack. As, we have observed that Yoon et al. just consider the security problems in the password change phase of Ku and Chen's scheme and repaired that phase only. They again presented a modified scheme with same security parameters as it was with previous security parameters. Thus, the security pitfalls are still exists in Yoon et al.'s scheme.

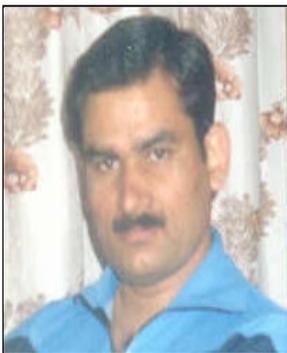

**Manoj Kumar** received the B.Sc. degree in mathematics from Meerut University Meerut, in 1993; the M. Sc. in Mathematics (Goldmedalist) from C.C.S.University Meerut, in 1995; the M.Phil. (Goldmedalist) in *Cryptography*, from Dr. B. R. A. University Agra, in 1996; the Ph.D. in *Cryptography*, in 2003. He also qualified the *National Eligibility Test* (NET), conducted by *Council of Scientific and Industrial Research* (CSIR), New Delhi- India, in 2000.

He also taught applied Mathematics at D. A. V. College, Muzaffarnagar, India from Sep 1999 to March 2001; at S.D. College of Engineering & Technology, Muzaffarnagar- U.P. – INDIA from March 2001 to Nov 2001; at Hindustan College of Science & Technology, Farah, Mathura- U.P. – INDIA, from Nov 2001 to March 2005. In 2004, the Higher Education Commission of U.P. has selected him.  Presently, he is working as lecturer in Department of Mathematics, R. K. Inter College Shamli- Muzaffarnagar- U.P. – INDIA.

He is a member of Indian Mathematical Society, Indian Society of Mathematics and Mathematical Science, Ramanujan Mathematical society, and Cryptography Research Society of India. His current research interests include Cryptography, Numerical analysis, Pure and Applied Mathematics.